\documentclass[superscriptaddress,twocolumn]{revtex4}
\usepackage[tbtags]{amsmath}
\usepackage{amsfonts}
\usepackage{mathrsfs}
\usepackage{amssymb}
\usepackage{graphicx}
\usepackage{epsfig,graphicx,times}
\usepackage{subfigure}
\usepackage{color}

\begin{document}

\title{Testing tripartite Mermin inequalities by spectral joint-measurements of qubits}

\author{J. S. Huang}
\affiliation{Quantum Optoelectronics Laboratory, School of Physics
and Technology, Southwest Jiaotong University, Chengdu 610031,
China} \affiliation{Centre for Quantum Technologies and Department
of Physics, National University of Singapore, 3 Science Drive 2,
Singapore 117542, Singapore}

\author{C. H. Oh\footnote{phyohch@nus.edu.sg}}
\affiliation{Centre for Quantum Technologies and Department of
Physics, National University of Singapore, 3 Science Drive 2,
Singapore 117542, Singapore}

\author{L. F. Wei\footnote{weilianfu@gmail.com}}
\affiliation{Quantum Optoelectronics Laboratory, School of Physics
and Technology, Southwest Jiaotong University, Chengdu 610031,
China} \affiliation{State Key Laboratory of Optoelectronic Materials
and Technologies, School of Physics and Engineering, Sun Yat-Sen
University Guangzhou 510275, China}
\begin{abstract}

It is well known that Bell
 inequality supporting the local
realism can be violated in quantum mechanics. Numerous tests of such
a violation have been demonstrated with bipartite entanglements.
Using spectral joint-measurements of the qubits, we here propose a
scheme to test the tripartite Mermin inequality (a three-qubit
Bell-type inequality) with three qubits dispersively-coupled to a
driven cavity. First, we show how to generate a three-qubit
Greenberger-Horne-Zeilinger (GHZ) state by only one-step quantum
operation. Then, spectral joint-measurements are introduced to
directly confirm such a tripartite entanglement. Assisted by a
series of single-qubit operations, these measurements are further
utilized to test the Mermin inequality. The feasibility of the
proposal is robustly demonstrated by the present numerical
experiments.

PACS number(s): 03.65.Ud, 42.50.Dv, 42.50.Pq

\end{abstract}

\maketitle

\section{Introduction}

Entanglement~\cite{Einstein} is at the heart of the quantum theory
and the crucial resource of quantum information
processing~\cite{Bouwmeester,Nielsen}. It is also one of the most
important ingredients of various intriguing phenomena, e.g., quantum
teleportation~\cite{Bennett,Bouwmeester2}, secret
sharing~\cite{Hillery}, and remote state
preparation~\cite{Nielsen2}, etc. Therefore, generating and
verifying the existence of entanglements are of great importance.

Since Bell inequality~\cite{Bell} and its CHSH
version~\cite{Clauser} was formulated to test the correlations
between two particles, numerous experiments with bipartite
entanglement, e.g., photons~\cite{Aspect}, trapped
ions~\cite{Rowe,Matsukevich}, neutrons~\cite{Hasegawa} and Josepson
junctions~\cite{Ansmann,WeiB}, etc., have been demonstrated to probe
the nonlocal nature of quantum mechanics. As all these experiments
support quantum mechanics and rule out the local hidden-variable
theories, Bell inequality can be served as an important witness of
quantum entanglement.

With the developments of quantum technology, entanglement shared by
multiple particles play more and more important roles for
large-scale quantum information processing and many-body quantum
mechanics. Experimentally, multipartite entangled states have been
demonstrated with photons~\cite{WPan,Resch,Pan,Zhao}, trapped
ions~\cite{Turchette,Sackett,Kirchmair}, Rydberg
atoms~\cite{Rauschenbeutel}, and also Josephson
circuits~\cite{Plastina,Wei}, etc. Basically, multipartite
entanglement can be robustly verified by the standard quantum-state
tomographic technique, i.e., reconstructing their density matrixes
by a series of quantum measurements. Instead, one can also verify
entanglement by testing the violation of the multipartite Bell-type
inequality, such as Mermin inequality~\cite{Mermin}:
\begin{eqnarray}
Q&=&|E(\theta'_1, \theta_2, \theta_3)+E(\theta_1, \theta'_2,
\theta_3) +E(\theta_1, \theta_2, \theta'_3)\nonumber\\
&&-E(\theta'_1, \theta'_2, \theta'_3)|\leq2
\end{eqnarray}
with three-qubit systems.
Indeed, the violation of this inequality has been experimentally
demonstrated with three-photon entanglement~\cite{Resch,Pan}. Above,
$\{\theta_1,$ $ \theta_2,$ $ \theta_3,$ $ \theta'_1,$ $ \theta'_2,$
$ \theta'_3\}$ are the set of controllable local-variables of the
three independent particles, and the correlation function
$E(\theta_1, \theta_2, \theta_3)$ is the ensemble average over the
measurement outcomes for the local settings: $\theta_1, \theta_2,
\theta_3$.

As a possible experimental demonstration, in this paper we discuss
how to perform the test of a tripartite Mermin inequality with three
qubits coupled to a driven cavity. Two main contributions in the
present proposal are: (i) an one-step approach is proposed to
generate the desired Greenberger-Horne-Zeilinger (GHZ)
state~\cite{GHZ}, and (ii) a spectral measurement method is
introduced to implement the joint measurements of these three
qubits. In principle, our proposal could be further generalized to
the cases with more than three particles.
 The paper is organized as:
In Sec. II, we briefly describe how to generate the desired
tripartite GHZ entangled state of three qubits coupled dispersively
to a driven cavity. Then, by introducing a spectral
joint-measurement method via detecting the photon transmission
through the driven cavity, we propose a simple two-step method to
confirm such a tripartite GHZ entanglement. In Sec. III, we propose
how to encode various local variables into the prepared GHZ
entanglement via performing suitable single-qubit operations, and
implement the test of the desired Mermin inequality by the
introduced joint-measurements. Discussions on the feasibility of our
proposal are given in Sec. IV.

\section{Generation and confirmation of the GHZ state of qubits coupled to a driven cavity-QED system}

\subsection{Preparation of tripartite GHZ state by only one-step quantum operation}

We consider a driven cavity-qubit system, wherein three qubits
without interbit interaction are respectively coupled to a common
driven cavity. In principle, such a cavity-qubit system can be
described by the Tavis-Cummings Hamiltonian ~\cite{Tavis} ($\hbar=1$
throughout the paper)
\begin{eqnarray}
 H_{\rm{TC}}=\omega_r\hat{a}^\dagger
 \hat{a}+\sum\limits_{j=1,2,3}[\frac{\omega_j}{2}\sigma_{z_j}+g_j(a^\dagger\sigma_{-_j}+a\sigma_{+_j})],
\end{eqnarray}
where $a^{(\dagger)}$ and $\sigma_{\pm_ j}$ are the ladder operators
for the photon field and the $j$th qubit, respectively; $\omega_r$
is the cavity frequency, $\omega_{j}$ the $j$th qubit transition
frequency, and $g_j$ the coupling strength between the $j$th qubit
and the cavity. The driving of the cavity can be modeled by
\begin{eqnarray}
 H_{d}= \varepsilon(t)(a^\dagger e^{-i\omega_d t}+a e^{i\omega_d t}),
\end{eqnarray}
where $\varepsilon(t)$ is the amplitude and $\omega_d$ the frequency
of the external drive.

Following Ref.~\cite{Blais}, after a displacement transformation
$D(\alpha)=\exp(\alpha\hat{a}^\dagger-\alpha^*\hat{a})$, the
displaced Hamiltonian of the composite system reads
\begin{eqnarray}
{H}_{\rm T}&=&D^\dagger(\alpha)(H_{\rm{TC}}+H_{d})
D(\alpha)-iD^\dagger(\alpha)\dot{D}(\alpha),\nonumber\\&=&
\omega_r{a}^\dagger{a}
+\sum\limits_{j=1,2,3}[\frac{\omega_j}{2}\sigma_{z_j}
+g_j(a^\dagger\sigma_{-_j}+a\sigma_{+_j})\nonumber\\&&+g_j(\alpha^*\sigma_{-_j}+\alpha\sigma_{+_j})].
\end{eqnarray}
Now, we let $\omega_d=\omega_j$, $\dot{\alpha}=-i\omega_r
\alpha-i\varepsilon\exp(-i\omega_dt)$, and work in a rotating frame
defined by
$\hat{U}_1=\exp[-it(\omega_{r}a^{\dagger}a+\sum\limits_{j=1,2,3}\omega_d\sigma_{z_j}/2)]$,
the effective Hamiltonian of the cavity-qubit system takes the form
\begin{eqnarray}
 \tilde{H}_{\rm T}
 &=&\sum_{j=1,2,3}[\Omega_{j}\sigma_{x_j}
+g_j(a^\dagger\sigma_{-_j}\exp{(-i\delta
 t)}\nonumber\\
 &&+a\sigma_{+_j}\exp{(i\delta t)})],
\end{eqnarray}
with the qubit-drive detuning $\delta=\omega_d-\omega_r$ and the
Rabi frequency $\Omega_{j}=\varepsilon g_j/\delta$. Changing to the
orthogonal bases
$|\pm_j\rangle=(|1_j\rangle\pm|0_j\rangle)/\sqrt{2}$,  and in the
interaction picture, we get
\begin{eqnarray}
&&H_{I}=
 \sum_{j=1,2,3}\frac{g_j}{2}a^\dagger \exp(-i\delta t)\left[|+_j\rangle \langle+_j|-|-_j\rangle
 \langle-_j|\right.\nonumber\\
 &&\left.+\exp{(i2\Omega_jt)}|+_j\rangle \langle-_j|- \exp{(-i2\Omega_jt)}|-_j\rangle
 \langle+_j|\right]+h.c.,\nonumber\\
\end{eqnarray}
where $|\pm_j\rangle$ are the eigenstates of operator $\sigma_{x_j}$
with eigenvalues $\pm1$. In the strong driving regime:
$\Omega\gg\delta,g$, we can eliminate the fast-oscillating terms in
Eq.~(6) and then have~\cite{Wu2,Zheng}
\begin{eqnarray}
H_{I}=
 \sum_{j=1,2,3}\frac{g_j}{2}\sigma_{x_j}
[a^\dagger\exp{(-i\delta t)}+a\exp{(i\delta t)}].
\end{eqnarray}

Note that the operator set $\{\sigma_{x_j}\sigma_{x_j'},
a^\dagger\sigma_{x_j},a\sigma_{x_j},1\}$ $(j,j'=1,2,3,$ and $j\neq
j')$ form a closed Lie algebra, the time evolution operator related
to the above Hamiltonian can be formally written as~\cite{JWei}
\begin{eqnarray}
U_I(t)&=&\exp{[-iC(t)]}\prod\limits_{j}\exp{[-i(B_{j}(t)a\sigma_{x_j}
+B^*_{j}(t)a^\dagger\sigma_{x_j})]}\nonumber\\
&\times&\prod\limits_{j\neq
j'}\exp{[-iA_{jj'}(t)\sigma_{x_j}\sigma_{x_{j'}}]},
\end{eqnarray}
with the parameters determined by
\begin{eqnarray}
&&A_{jj'}(t)=
\frac{g_jg_{j'}}{4\delta}[\frac{1}{i\delta}(\exp{(-i\delta
t)}-1)+t],\nonumber\\
&&B_{j}(t)=\frac{g_j}{i2\delta}[\exp{(i\delta t)}-1],\nonumber\\
&&C(t)=\sum\limits_j\frac{g_j^2}{4\delta}[\frac{1}{i\delta}(\exp{(-i\delta
t)}-1)+t],
\end{eqnarray}
and $A_{jj'}(0)=B_j(0)=C(0)=0$.

Suppose that all the qubit-cavity couplings are homogeneous, i.e.,
$g_j=g$ (for $j=1,2,3$) and set $ \delta t=2n\pi$ for integer $n$,
we have $B(t)=B^*(t)=0$. Then, the time evolution operator reduces
to a simple form
\begin{eqnarray}
U_I(t)= \exp{(-i\frac{g^2}{\delta}tS_{x}^2)},
\end{eqnarray}
with $S_x=\sum_{j=1}^3\sigma_{x_j}/2$. Return to the Schr\"{o}inger
picture,
\begin{eqnarray}
 U_S(t)&=&U_0(t)U_I(t)\nonumber\\
&=& \exp{(-i\omega a^\dagger at)}\prod_j
 \exp{(-i\Omega_j\sigma_{x_j}t)}U_I(t)\nonumber\\
 &=& \exp{(-i\omega a^\dagger at)}
 \exp{(-i2\Omega S_{x}t-i\frac{g^2}{\delta}tS_{x}^2)}.
\end{eqnarray}
Here $\Omega_j=\Omega$ for $g_j=g$ mentioned above. Note that the
effective coupling $S_{x}^2$ can be utilized to directly realize the
multi-qubit GHZ state, once the relevant parameters are
appropriately chosen~\cite{Mo}. Assume that the three-qubit register
is initially at the state
\begin{eqnarray}
|\psi(0)\rangle=|000\rangle,
\end{eqnarray}
where $|1\rangle$ ($|0\rangle$) denotes the eigenstate of
$\sigma_z$, $\sigma_z |1\rangle=1$, $\sigma_z |0\rangle=-1$. Using
the spin representation of atomic states for the operator $S_z$,
($S_z=\sum_{j=1}^3\sigma_{z_j}/2$), the three-qubit states
$|000\rangle$ and $|111\rangle$ can be expressed as collective
states $|3/2,-3/2\rangle$ and $|3/2,3/2\rangle$, respectively. Here,
$|J=3/2, M\rangle$ is the eigenstate of the operators $S_z$ with the
eigenvalue $M$, $M=-J,...,J$. In terms of the eigenstates of
$S_x$~\cite{Mo}, we have
\begin{eqnarray}
|3/2,-3/2\rangle=\sum_{M=-3/2}^{3/2}c_M|3/2,M\rangle_x ,
\end{eqnarray}
and
\begin{eqnarray}
|3/2,3/2\rangle=\sum_{M=-3/2}^{3/2}c_M(-1)^{3/2-M}|3/2,M\rangle_x,
\end{eqnarray}
where $M=M'+1/2$ and $M'$ is an integer. As a consequence, the
evolution of the system can be conveniently expressed as (up to a
global phase factor)
\begin{eqnarray}
|\psi(t)\rangle&=&U_S(t)|\psi(0)\rangle\nonumber\\
&=&\sum_{M=-3/2}^{3/2}c_M\exp{[-i2\Omega t M-i\frac{g^2}{\delta}t
M^2]}|3/2,M\rangle_x
\nonumber\\
&=&\frac{1}{\sqrt{2}}(|3/2,-3/2\rangle+ i|3/2,3/2\rangle),
\end{eqnarray}
with the choice $g^2t/\delta=(4k+1)\pi/2$ and $\Omega t=(2m+3/4)\pi$
($k$, $m$ are integers). Obviously, at $t=T_n$ the desired GHZ
state~\cite{Mo,Zheng,Wang}
\begin{eqnarray}
|\psi(T_n)\rangle=U_S(T_n)|\psi(0)\rangle=
\frac{1}{\sqrt{2}}(|000\rangle+ i|111\rangle)
\end{eqnarray}
is obtained. In the above, the relations of the integers $k$, $m$
and $n$ are given by
\begin{eqnarray}
 n=\frac{\delta^2}{g^2}k+\frac{\delta^2}{4g^2},\,\,
 n=\frac{2\delta^2}{\varepsilon g}m+\frac{3\delta^2}{4\varepsilon
 g}.
\end{eqnarray}

\subsection{Confirming the existence of the GHZ entanglement}

The GHZ state prepared above by one-step operation can be robustly
confirmed by using the standard quantum-state tomographic technique,
i.e., reconstructing its density matrix. Such an approach was
usually utilized to confirm the quantum state engineering in
trapped-ions~\cite{Leibfried}, linear optics~\cite{Resch,Pan,Zhao}
and the solid-state qubits~\cite{Liu,Neeley,DiCarlo}, etc. However,
these confirmations require many kinds of {\it{single-basis}}
projective measurements assisted by a series of quantum operations,
and thus $2^N-1$ kinds of projections are needed for reconstructing
a $N\times N$-matrix, in principle.

Fortunately, a significantly simple approach, i.e., spectral
joint-measurements of the qubits~\cite{arXiv,Huang}, can be utilized
to high-effectively implement the desired confirmation. By this
approach the states of the qubits could be jointly detected by
probing the steady-state transmission spectra of the driven cavity,
which is commonly coupled to the qubits. For the present case, the
qubit-cavity detuning $\Delta=\omega-\omega_r$ is assumed to be much
larger than the coupling $g$ (i.e., the system works in the
dispersive regime) and the qubit-cavity couplings take the form
$H_c=a^\dagger a\sum_{j=1}^3\Gamma_j\sigma_{z_j}$. This indicates
that the qubits cause the state-dependent frequency shift of the
cavity. For example, if the qubits is prepared at the joint
eigenstate $|000\rangle$ (or $|111\rangle$) of the three qubits,
then the frequency of the cavity is shifted as
$-\sum_{j=1}^3\Gamma_j$ (or $\sum_{j=1}^3\Gamma_j$). Due to such a
pull, the frequency of the transmitted photons through the cavity is
shifted, which is dependent on the {\it joint} eigenstate of the
qubits. Thus, the steady-state transmission spectra through the
driven cavity can mark all the possible joint eigenstates of the
qubits. Generally, unknown qubits should be denoted as a
superposition of all the possible joint eigenstates of the qubits.
As a consequence, the measured transmission spectra $\langle
a^{\dagger}a(\omega_d)\rangle_{\rm{ss}}$ of the driven cavity may
appear multiple peaks versus the driving frequency $\omega_d$ (see
the Appendix for the detailed derivation); each peak marks one of
the possible joint eigenstates of the qubits, and its relative
height corresponds to the probability of this state superposed in
the three-qubit unknown state.

\begin{figure}[htbp]
  \centering
  \includegraphics[width=8cm]{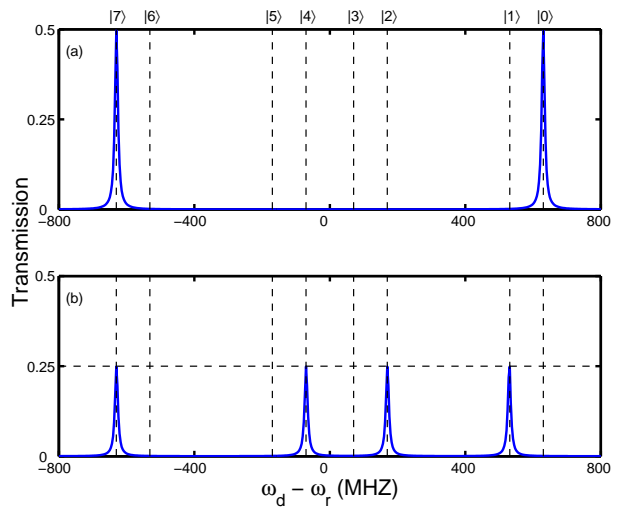}\\
  \caption {(Color online) Two spectral joint-measurements to confirm the GHZ entanglement:
  (a) directly for the GHZ state: $(|000\rangle+ |111\rangle)/{\sqrt{2}}$, and
(b) for the state
$(|000\rangle+|011\rangle+|101\rangle+|110\rangle)/2$ generated
after performing unitary operations on the GHZ state. Here,
  the parameters are selected as $(\Gamma_1, \Gamma_2, \Gamma_3, \kappa)
  =2\pi\times(50, 230, 350, 1.69)$MHz, and $|0\rangle = |000\rangle,
  |1\rangle = |001\rangle,
  |2\rangle = |010\rangle, |3\rangle = |011\rangle,
   |4\rangle = |100\rangle,
  |5\rangle = |101\rangle, |6\rangle = |110\rangle$, and
  $|7\rangle = |111\rangle$, respectively.}
\end{figure}

Specifically, for the GHZ state prepared above the steady-state
transmission spectra of the driven cavity should reveal a two-peak
structure, see, e.g., Fig.~1(a) with the typical parameters:
$(\Gamma_1, \Gamma_2, \Gamma_3, \kappa)=2\pi\times(50, 230, 350,
1.69)$MHz. To show clearly the simulated results for the testing,
the parameters $\Gamma_j$ could be adjusted by adiabatically tuning
the qubit-transition frequencies. Desirably, the frequency-shift
locations: $-\Gamma_1-\Gamma_2-\Gamma_3$,
$\Gamma_1+\Gamma_2+\Gamma_3$ and the relative heights of these two
peaks: $0.5$, $0.5$, indicate that two joint eigenstates,
$|000\rangle$ and $|111\rangle$, are superposed in the measured
state with the same superposition probability $0.5$. Of course, such
a spectral joint-measurement result is just a necessary but not
sufficient condition to assure the desired GHZ state, since a
statistical mixture of these two joint eigenstates may also yield
the same spectral distributions. To confirm the state
$|\psi_{\rm{GHZ}}\rangle$ is indeed the coherent superposition of
the states $|000\rangle$ and $|111\rangle$, we need another spectral
joint-measurement by using the quantum coherent effect. This can be
achieved by first applying the unitary operation
$\prod_{j=1}^3R_{y_j}(\pi/4)=\prod_{j=1}^3\exp{(i\pi\sigma_{y_j}/4)}$
to each qubit, yielding the evolution
\begin{eqnarray}
|\psi_{\rm{GHZ}}\rangle&\rightarrow&|\psi'_{\rm{GHZ}}\rangle=R_{y_1}(\frac{\pi}{4})R_{y_2}(\frac{\pi}{4})R_{y_3}(\frac{\pi}{4})|\psi_{\rm{GHZ}}\rangle \nonumber\\
 &=&\frac{1}{2}(|000\rangle+|011\rangle+|101\rangle+|110\rangle),
\end{eqnarray}
and then performing the spectral joint-measurement. It is expected
~\cite{arXiv,Huang} that four peaks with the same relative height
$0.25$ should be observed (see, e.g., Fig.~1(b)), if the prepared
state is nothing but the desired tripartite GHZ state. However, if
the prepared state is a mixture of the joint eignestates
$|000\rangle$ and $|111\rangle$, then eight peaks would be observed.

\section{Testing tripartite Mermin inequality by spectral joint-measurements}

With the GHZ state, we now discuss how to test the tripartite Mermin
inequality (1) by jointly measuring the three qubits, simultaneously
coupled to a driven cavity. The test includes the following two
steps.

First, local parameters $\theta_j (j=1,2,3)$ are encoded into the
generated GHZ state (16) by performing the single-qubit
Hadamard-like operations
\begin{eqnarray}
R_j(\theta_j)&=&R_{z_j}(\theta_j/2)R_{x_j}(\pi/4)R_{z_j}(-\theta_j/2)\nonumber\\
&=&\frac{1}{\sqrt{2}}\left(
\begin{array}{cc}
1 & ie^{i\theta_j} \\
ie^{-i\theta_j} & 1 \\
\end{array}\right).
\end{eqnarray}
Here, the typical single-qubit gates
$R_{z_j}(\theta)=\exp{(i\sigma_{z_j}\theta)}$ and
$R_{x_j}(\theta)=\exp{(i\sigma_{x_j}\theta)}$ can be
relatively-easily implemented, see e.g.~\cite{Blais,Huang}. After
these encoding operations, the generated GHZ state
$|\psi_{\rm{GHZ}}\rangle$ is changed as
\begin{eqnarray}
|\psi{''}_{\rm{GHZ}}\rangle&=&R_1(\theta_1)R_2(\theta_2)R_3(\theta_3)|\psi_{\rm{GHZ}}\rangle\nonumber\\
&=&\frac{1}{4}[(1+e^{i(\theta_1+\theta_2+\theta_3)})|000\rangle
\nonumber\\&&+(ie^{-i\theta_3}-ie^{i(\theta_1+\theta_2)})|001\rangle
\nonumber\\&&+(ie^{-i\theta_2}-ie^{i(\theta_1+\theta_3)})|010\rangle
\nonumber\\&&+(-e^{-i(\theta_2+\theta_3)}-e^{i\theta_1})|011\rangle
\nonumber\\&&+(ie^{-i\theta_1}-ie^{i(\theta_2+\theta_3)})|100\rangle
\nonumber\\&&+(-e^{-i(\theta_1+\theta_3)}-e^{i\theta_2})|101\rangle
\nonumber\\&&+(-e^{-i(\theta_1+\theta_2)}-e^{i\theta_3})|110\rangle
\nonumber\\&&+(i-ie^{-i(\theta_1+\theta_2+\theta_3)})|111\rangle].
\end{eqnarray}
Second, we perform the joint projective-measurements to determine
the required correlation functions $E(\theta_1, \theta_2, \theta_3)$
for various combinations of these local variables.

Experimentally, the above two steps can be repeated many times, and
thus the correlation function can be determined by
\begin{eqnarray}
E(\theta_1, \theta_2,
\theta_3)&=&P_{111}+P_{100}+P_{010}+P_{001}\nonumber\\
&-&P_{011}-P_{101}-P_{110}-P_{000}.
\end{eqnarray}
Here, $\sum_{i,j,k=0,1}P_{ijk}=1$ with $P_{ijk}$ is the probability
of the state $|\psi''_{\rm GHZ}\rangle$ collapsing to the joint
basis $|ijk\rangle$. With these projective measurements, various
correlation functions required can be measured and then the
tripartite Mermin inequality (1) can be tested. Theoretically, the
correlation function can be easily calculated as
\begin{eqnarray}
E(\theta_1,
\theta_2,\theta_3)&=&\langle\psi{''}_{\rm{GHZ}}|\hat{P}_T
|\psi{''}_{\rm{GHZ}}\rangle
\nonumber\\&=&-\cos(\theta_1+\theta_2+\theta_3),
\end{eqnarray}
with the joint projective operator
$\hat{P}_T=\sigma_{z_1}\otimes\sigma_{z_2}\otimes\sigma_{z_3}=|111\rangle\langle
111|+|100\rangle\langle 100|+|010\rangle\langle
010|+|001\rangle\langle 001|-|011\rangle\langle
011|-|101\rangle\langle 101|-|110\rangle\langle
110|-|000\rangle\langle 000|$. For the suitable choices of the local
observables, e.g., $\{\theta_1, \theta_2,$ $\theta_3, \theta'_1,$ $
\theta'_2, \theta'_3\}$ $=\{0,$ $\pi/4, \pi/2,$ $\pi/4,$ $\pi/4,$
$\pi\}$, we have the ideal value of the $Q$-parameter in Eq.~(1):
\begin{eqnarray}
Q_i=\sqrt{2}+1>2.
\end{eqnarray}
This indicates that the inequality (1), namely $Q_i\leq2$, is
violated. Furthermore, for the parameters $\{\theta_1, \theta_2,
\theta_3, \theta'_1, \theta'_2, \theta'_3\}=\{\pi/4, 0, 0,3\pi/4,
\pi/2,\pi/2 \}$, the above tripartite Mermin inequality is maximally
violated, i.e., $Q_i=2\sqrt{2}$.

\begin{figure}[htbp]
  \centering
  \includegraphics[width=8cm]{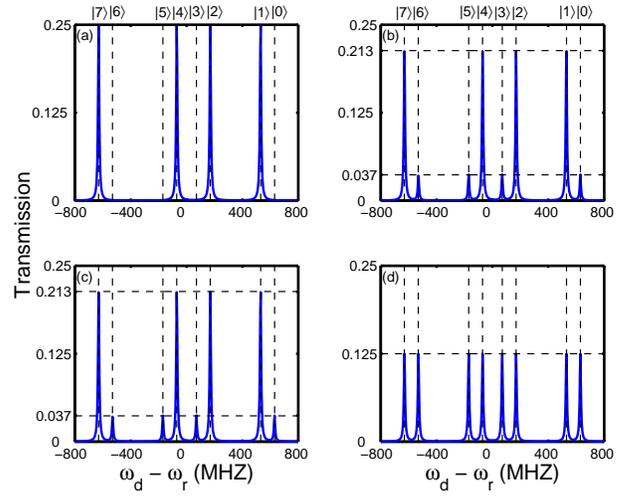}\\
  \caption {(Color online) Transmission spectra of the driven cavity versus the detuning
  for the evolved state $|\psi{''}_{\rm{GHZ}}\rangle$ with
  the classical variables
  $\{\theta_1, \theta_2, \theta_3, \theta'_1, \theta'_2,
\theta'_3\}=\{0, \pi/4, \pi/2, \pi/4, \pi/4, \pi\}$. Here, (a)-(d)
correspond respectively to the parameters
   $\{\theta'_1,\theta_2,\theta_3\}$,
   $\{\theta_1,\theta'_2,\theta_3\}$,
   $\{\theta_1,\theta_2,\theta'_3\}$, and $\{\theta'_1,\theta'_2,\theta'_3\}$.
With these spectral distributions, the correlation functions
required for testing the Mermin inequality (1) can be calculated.
Other parameters of the system are the same as those used in
Fig.~1.}
\end{figure}

Like in the usual tomographic reconstructions only one basis, e.g.,
$|ijk\rangle$, is collapsed for one kind of projective measurement
$\hat{P}_{ijk}=|ijk\rangle\langle ijk|$. This implies that seven
kinds of projective measurements are required to complete the above
joint projection $\hat{P}_T$. However, by the spectral
joint-measurements introduced in Refs.~\cite{arXiv,Huang}, the
probabilities $P_{ijk} (i,j,k=0,1)$ can be determined simultaneously
by just the spectral measurements of the transmission through the
driven cavity; each peak of the transmission spectra marks one of
the basis $|ijk\rangle$, and its relative height refers to the
relevant probability $P_{ijk}$. Specifically, for one set of
classical variables $\{\theta_1,$ $\theta_2, \theta_3,$ $\theta'_1,$
$\theta'_2,$ $\theta'_3\}$ $=\{0,$ $\pi/4, \pi/2,$ $\pi/4,$ $\pi/4,$
$\pi\}$, Figs.~2(a-d) show how the spectra of the driven cavity
distribute (versus the qubit-driving detuning) for the state (20).
For instance, four peaks, marking respectively the basis states
$|000\rangle$, $|011\rangle$, $|101\rangle$, and $|110\rangle$, are
shown in Fig.~2(a). Their relative heights are equivalent:
$P_{000}=P_{011}=P_{101}=P_{110}=0.25$. Thus, the correlation
function between three local variables can be easily calculated as
\begin{eqnarray}
\left\{ \begin{array}{cc}
E(\pi/4,0,0)=1,\nonumber\\
E(\pi/2,0,0)= 0.704, \nonumber\\
E(\pi/4,\pi/2,0)=0.704,\nonumber\\
E(\pi/2,\pi/2,0)=0.
\end{array}
\right.
\end{eqnarray}
Consequently, the numerical experimental result of the $Q$-parameter
is
\begin{eqnarray}
Q_e=2.408\approx \sqrt{2}+1>2,
\end{eqnarray}
and thus the tripartite Mermin inequality is violated. Similarly,
for another set of classical variables $\{\theta_1,$ $ \theta_2,$
$\theta_3, \theta'_1,$ $\theta'_2, \theta'_3\}=\{\pi/4, 0, 0,3\pi/4,
\pi/2,\pi/2 \}$, Figs.~3(a-d) show all the probabilities of eight
bases in the present three-qubit system. Again, the involved
correlation functions are calculated as
$(E(\theta'_1,\theta_2,\theta_3)$, $E(\theta_1,\theta'_2,\theta_3)$,
$E(\theta_1,\theta_2,\theta'_3)$,
$E(\theta'_1,\theta'_2,\theta'_3))$=(0.704, 0.704, 0.704, -0.704).
As a consequence,
\begin{eqnarray}
Q_e=2.816\approx 2\sqrt{2}>2,
\end{eqnarray}
and the Mermin inequality (1) is violated more strongly.

\begin{figure}[htbp]
  \centering
  \includegraphics[width=8cm]{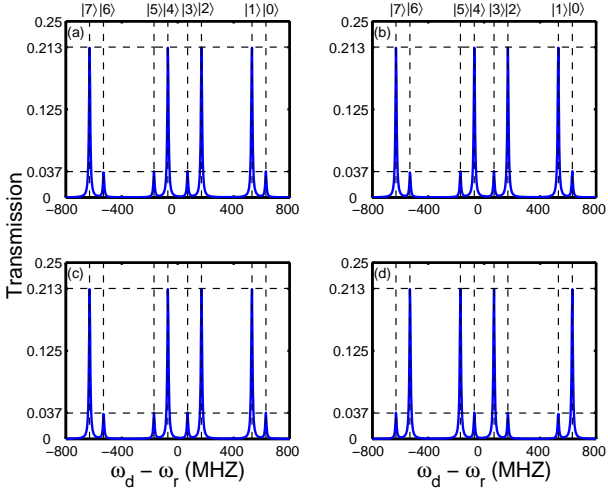}\\
  \caption {(Color online) Transmission spectra of the driven cavity versus
  the qubit-drive detuning for the set
  of local variables
  $\{\theta_1, \theta_2, \theta_3, \theta'_1, \theta'_2,
\theta'_3\}=\{\pi/4, 0, 0,3\pi/4, \pi/2,\pi/2 \}$.
  Other parameters are the same as those in Fig.~1.}
\end{figure}

\section{Discussion}

We have proposed a direct and experimentally-feasible scheme to test
tripartite Mermin inequality with cavity-qubit system, wherein
quantum state of three qubits without any direct interbit coupling
is detected by measuring the dispersively-coupled cavity. We have
numerically demonstrated that the local-variable-dependent
probabilities of various bases superposed in the
local-variable-encoded GHZ state can be directly read out by the
cavity transmission. With these probabilities, various correlation
functions on the local variables of individual qubits are easily
calculated, and consequently the violations of the three-particle
Mermin inequality are tested. Specifically, a few examples were
utilized to numerically confirm the tests. Certainly, the present
proposal could be generalized to test various Bell-type inequalities
with more than three qubits in a straightforward way.

Note that in our numerical-experiments little deviations exist
between our estimated results and the ideal predictions. For
example, if the local variables are set as $\{\theta_1, \theta_2,$
$\theta_3, \theta'_1,$ $ \theta'_2, \theta'_3\}=\{0, \pi/4, \pi/2,
\pi/4, \pi/4, \pi\}$, the values of $Q$-paramter given by our
numerical experiments is $Q_{e}=2.408$, which deviates the ideal
values $Q_{i}=\sqrt{2}+1$ with a quantity $\Delta Q=Q_i-Q_e=0.006$.
Also, for local variables $\{\theta_1, \theta_2, \theta_3,
\theta'_1, \theta'_2, \theta'_3\}=\{\pi/4, 0, 0,3\pi/4, \pi/2,\pi/2
\}$ the inequality (1) should be maximally violated with
$Q_i=2\sqrt{2}$, but our numerical experiment yields
$Q_e=2.816=Q_i-0.012$. These deviations are due to the existence of
the dissipation of the cavity, which yields various finite widths of
the transmission spectra through the driven cavity. As a
consequence, the relative heights of the measured peaks are lower
than those of the ideal $\delta$-type peaks. Therefore, the
violations of the Mermin inequalities are less than those for the
ideal cases. But, it is sufficient to show the violation of the
Mermin inequality.

\section*{Acknowledgments}

This work was supported in part by the National Science Foundation
grant No. 10874142, 90921010, and the National Fundamental Research
Program of China through Grant No. 2010CB92304, and the Fundamental
Research Funds for the Central Universities No. SWJTU09CX078, and
Centre for quantum technology grant number: WBS: R-710-000-008-271.

\section*{APPENDIX: TRANSMISSION OF A DRIVEN CAVITY}
\makeatletter
\renewcommand\theequation{A\@arabic\c@equation }
\makeatother \setcounter{equation}{0}

In this appendix, the transmission spectrum of a three-qubit in a
driven cavity is calculated in detail. The transition frequencies of
the three qubits are denoted as $\omega_1$, $\omega_2$ and
$\omega_3$, respectively. We assume that the dispersive condition
\begin{equation}
0<\frac{g_j}{\Delta_j},\,\frac{g_jg_{j'}}{\Delta_j\Delta_{jj'}},\,\frac{g_jg_{j'}}{\Delta_{j'}\Delta_{jj'}}\ll
1,\,\,j\neq j'=1,2,3,
\end{equation}
is satisfied for ensuring the effective dispersive coupling
$\sigma_{z_j}\hat{a}^\dagger\hat{a}$ between the $j$th qubit and the
cavity. These conditions also ensure that the interbit interactions
are negligible. Also, $\Delta_j=\omega_r-\omega_j$ denotes the
detuning between the $j$th qubit and the cavity, and
$\Delta_{jj'}=\omega_j-\omega_j'$ the detuning between the $j$th and
$j'$th qubits.

In a framework rotating at $\omega_{d}$, the effective Hamiltonian
of the
 qubit-cavity system is
\begin{eqnarray}
\tilde{H}&=&(-\delta+\Gamma_1\sigma_{z_1}+\Gamma_2\sigma_{z_2}+\Gamma_3\sigma_{z_3})\hat{a}^\dagger\hat{a}\nonumber\\
&+&\frac{\tilde{\omega}_1}{2}\sigma_{z_1}+\frac{\tilde{\omega}_2}{2}\sigma_{z_2}+\frac{\tilde{\omega}_3}{2}\sigma_{z_3}
+\epsilon(\hat{a}^\dagger+\hat{a}),
\end{eqnarray}
where $\Gamma_j=g_j^2/\Delta_{j}$,
$\tilde{\omega}_j=\omega_j+\Gamma_j$, $(j=1,2,3)$,
 and $\delta=\omega_d-\omega_r$ is the detuning of the cavity
from the driving. The master equation for the complete system reads
\begin{eqnarray}
\dot{\varrho}&=&-i[\tilde{H},\varrho]
+\kappa(\hat{a}\varrho\hat{a}^\dagger-\hat{a}^\dagger\hat{a}\varrho/2-\varrho\hat{a}^\dagger\hat{a}/2),
\end{eqnarray}
where $\varrho$ is the density matrix of the qubit-cavity system.

From the above master equation, the equations of motion for the mean
values of various expectable operators are
\begin{subequations}
\label{eq:whole}
\begin{eqnarray}
\frac{d\langle\hat{a}^\dagger\hat{a}\rangle}{dt}&
=&-\kappa{\langle\hat{a}^\dagger\hat{a}\rangle}-2\epsilon\mathrm{Im}{\langle\hat{a}\rangle},\label{subeq:1}
\end{eqnarray}
\begin{eqnarray}
\frac{d\langle\hat{a}\rangle}{dt}&
=&(i\delta-\frac{\kappa}{2}){\langle\hat{a}\rangle}-i\epsilon\nonumber\\
&-&i\Gamma_1{\langle\hat{a}\sigma_{z_1}\rangle}-i\Gamma_{2}{\langle\hat{a}\sigma_{z_2}\rangle}-i\Gamma_{3}{\langle\hat{a}\sigma_{z_3}\rangle},\label{subeq:2}
\end{eqnarray}
with
\begin{eqnarray}
\frac{d\langle\hat{a}\sigma_{z_1}\rangle}{dt}&=&(i\delta-\frac{\kappa}{2}){\langle\hat{a}\sigma_{z_1}\rangle}
-i\epsilon{\langle\sigma_{z_1}\rangle}
-i\Gamma_2{\langle\hat{a}\sigma_{z_1}\sigma_{z_2}\rangle}\nonumber\\&-&i\Gamma_1{\langle\hat{a}\rangle}-i\Gamma_3{\langle\hat{a}\sigma_{z_1}\sigma_{z_3}\rangle},\label{subeq:3}
\end{eqnarray}
\begin{eqnarray}
\frac{d\langle\hat{a}\sigma_{z_2}\rangle}{dt}&=&(i\delta-\frac{\kappa}{2}){\langle\hat{a}\sigma_{z_2}\rangle}
-i\epsilon{\langle\sigma_{z_2}\rangle}-i\Gamma_1{\langle\hat{a}\sigma_{z_2}\sigma_{z_1}\rangle}\nonumber\\&-&i\Gamma_2{\langle\hat{a}\rangle}-i\Gamma_3{\langle\hat{a}\sigma_{z_2}\sigma_{z_3}\rangle},\label{subeq:4}
\end{eqnarray}
\begin{eqnarray}
\frac{d\langle\hat{a}\sigma_{z_3}\rangle}{dt}&=&(i\delta-\frac{\kappa}{2}){\langle\hat{a}\sigma_{z_3}\rangle}-i\epsilon{\langle\sigma_{z_3}\rangle}
-i\Gamma_1{\langle\hat{a}\sigma_{z_3}\sigma_{z_1}\rangle}\nonumber\\&-&i\Gamma_3{\langle\hat{a}\rangle}-i\Gamma_2{\langle\hat{a}\sigma_{z_3}\sigma_{z_2}\rangle},\label{subeq:5}
\end{eqnarray}
and
\begin{eqnarray}
\frac{d\langle\hat{a}\sigma_{z_1}\sigma_{z_2}\rangle}{dt}&=&(i\delta-\frac{\kappa}{2}){\langle\hat{a}\sigma_{z_1}\sigma_{z_2}\rangle}
-i\epsilon{\langle\sigma_{z_1}\sigma_{z_2}\rangle} \nonumber\\&-&
i\Gamma_1{\langle\hat{a}\sigma_{z_2}\rangle}-i\Gamma_2{\langle\hat{a}\sigma_{z_1}\rangle}
\nonumber\\&-&i\Gamma_3{\langle\hat{a}\sigma_{z_1}\sigma_{z_2}\sigma_{z_3}\rangle},
\label{subeq:6}
\end{eqnarray}
\begin{eqnarray}
\frac{d\langle\hat{a}\sigma_{z_2}\sigma_{z_3}\rangle}{dt}&=&(i\delta-\frac{\kappa}{2}){\langle\hat{a}\sigma_{z_2}\sigma_{z_3}\rangle}
-i\epsilon{\langle\sigma_{z_2}\sigma_{z_3}\rangle} \nonumber\\&-&
i\Gamma_2{\langle\hat{a}\sigma_{z_3}\rangle}-i\Gamma_3{\langle\hat{a}\sigma_{z_2}\rangle}
\nonumber\\&-&i\Gamma_1{\langle\hat{a}\sigma_{z_1}\sigma_{z_2}\sigma_{z_3}\rangle},
\label{subeq:7}
\end{eqnarray}
\begin{eqnarray}
\frac{d\langle\hat{a}\sigma_{z_1}\sigma_{z_3}\rangle}{dt}&=&(i\delta-\frac{\kappa}{2}){\langle\hat{a}\sigma_{z_1}\sigma_{z_3}\rangle}
-i\epsilon{\langle\sigma_{z_1}\sigma_{z_3}\rangle} \nonumber\\&-&
i\Gamma_1{\langle\hat{a}\sigma_{z_3}\rangle}-i\Gamma_3{\langle\hat{a}\sigma_{z_1}\rangle}
\nonumber\\&-&i\Gamma_2{\langle\hat{a}\sigma_{z_1}\sigma_{z_2}\sigma_{z_3}\rangle},
\label{subeq:8}
\end{eqnarray}
\begin{eqnarray}
\frac{d\langle\hat{a}\sigma_{z_1}\sigma_{z_2}\sigma_{z_3}\rangle}{dt}&=&(i\delta-\frac{\kappa}{2}){\langle\hat{a}\sigma_{z_1}\sigma_{z_2}\sigma_{z_3}\rangle}
\nonumber\\&-&i\epsilon{\langle\sigma_{z_1}\sigma_{z_2}\sigma_{z_3}\rangle}
-i\Gamma_1{\langle\hat{a}\sigma_{z_2}\sigma_{z_3}\rangle}
\nonumber\\&-&i\Gamma_2{\langle\hat{a}\sigma_{z_1}\sigma_{z_3}\rangle}
-i\Gamma_3{\langle\hat{a}\sigma_{z_1}\sigma_{z_2}\rangle},\nonumber\\
\label{subeq:9}
\end{eqnarray}
\begin{eqnarray}
\frac{d\langle\sigma_{z_1}\rangle}{dt}=\frac{d\langle\sigma_{z_2}\rangle}{dt}=\frac{d\langle\sigma_{z_3}\rangle}{dt}
=0 ,\label{subeq:10}
\end{eqnarray}
\begin{eqnarray}
\frac{d\langle\sigma_{z_1}\sigma_{z_2}\rangle}{dt}=\frac{d\langle\sigma_{z_1}\sigma_{z_3}\rangle}{dt}=\frac{d\langle\sigma_{z_2}\sigma_{z_3}\rangle}{dt}=0
,\label{subeq:11}
\end{eqnarray}
\begin{eqnarray}
\frac{d\langle\sigma_{z_1}\sigma_{z_2}\sigma_{z_3}\rangle}{dt}=0,\label{subeq:12}
\end{eqnarray}
\end{subequations}
The steady-state distribution of the intracavity photon number can
be obtained by solving the Eqs.~(A4 a-i) under the steady-state
condition, i.e., all the derivatives in the left sides of above
equations equate $0$. Then, by numerical method, the steady-state
average photons number inside the cavity can be obtained. Similar to
the single-qubit and two-qubit cases in Ref.~\cite{arXiv,Huang},
information of these eight basis states in arbitrary three-qubit
state can be extracted from the spectra of the cavity transmission,
since each peak marks one of the eight bases, and its relative
height refers to its probability superposed in the measured
three-qubit state.

\end{document}